\documentclass[twocolumn,showpacs,showkeys,preprintnumbers,amsmath,amssymb,nofootinbib]{revtex4}
\usepackage{graphicx}
\usepackage{dcolumn}
\usepackage{bm}
\begin{document}
\title{\bf Gluon emission  in Quark-Gluon Plasma}
\author{S.  V.  Suryanarayana}
\email{snarayan@barc.gov.in}
\email{suryanarayan7@yahoo.com}
\affiliation{  Nuclear Physics Division, Bhabha Atomic Research Centre, Trombay, Mumbai
400 085, India}
\begin{abstract}{ 
Gluon radiation is an important mechanism for parton energy loss as the parton 
traverses the  quark gluon plasma (QGP) medium. We studied the gluon emission 
in QGP using AMY formalism. In the present work, we obtained gluon emission amplitude 
{\bf F}({\bf h},p,k) 
function , which is a solution of the  integral equations describing  gluon radiation 
 including   Landau-Pomeranchuk-Migdal (LPM) effects, using  iterations method.
We define a new dynamical scale for gluon emission denoted 
by $x$.  The gluon emission rate is obtained by integrating these amplitude 
function  over {\bf h}. We show that these  obey a simple scaling in terms 
of this dynamical  variable $x$.  We define the gluon emission function $g(x)$  
for gluon radiation for the three processes 
$g\rightarrow gg$,$q\rightarrow gq$ and $g\rightarrow q\bar{q}$. 
In terms of this $g(x)$ function, the parton energy loss 
calculations, due to medium induced gluon radiation,  may become  simplified.
}\end{abstract}
\maketitle
\par
\noindent

\begin{figure}
\includegraphics[height=5.0cm,width=8.0cm]{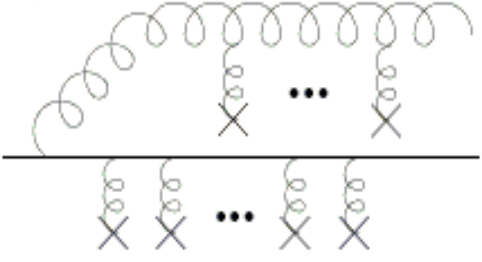}
\caption{gluon radiation processes  that contribute at order ${\alpha_s}$.}
\label{ladderg}
\end{figure}

Lattice quantum chromo dynamics (LQCD) calculations  \cite{karsch}  predict a transition 
from confined state in hadrons to  a deconfined state of quarks and gluons above a 
temperature of 170 MeV or an energy density above 1GeV/fm$^3$ .  In the  relativistic heavy ion
collisions at RHIC at BNL  with  an energy density above 5GeV/fm$^3$, experimental 
measurements of several observables indicate such a transition to  a deconfined state of 
matter \cite{brahms,phobos,star,phenix},  for details see the reviews  
\cite{emsig,qgpreview,rapp,enteria}.  It  is currently believed that 
this deconfined state   consists of  a strongly interacting Quark Gluon 
Plasma (sQGP), behaving nearly  like a perfect liquid \cite{qgpreview}. Among these 
observables, jet-quenching phenomenon is so far an important signal for a  
hot dense medium formed. Jet suppression has its origin in parton energy 
loss in the quark matter by gluon radiation, which distinctly differs from
energy loss in hadronic matter.   For example the suppression of high-$p_T$  
pions,  from 3GeV to 10GeV, of  BNL experiments can be explained by assuming 
a deconfined state.\\
We studied gluon radiation problem as this has direct application to the 
energy loss of partons while traversing the  QGP medium due to the gluon bremsstrahlung 
processes.  In addition to radiation, the elastic energy loss of partons traversing  the 
QGP medium is important for heavy quark quenching, observed in RHIC experiments.
For an exposition of  theoretical and experimental results on parton energy loss  
the readers may see an excellent review \cite{mazum}.\\
 In the  present work, we study the gluon radiation mechanism. The coherent radiation processes
 involve multiple scatterings of the partons in the QGP medium during the gluon formation time.
This leads to interference effects known as Landau-Pomeranchuk-Migdal effect (LPM). Gluon 
emission is discussed widely in literature \cite{baier}-\cite{moore1}.  These works treated 
the parton energy loss on the basis of avarage energy loss dependeing on the path length. 
As emphasized in \cite{moore2}, the bremsstrahlung (gluon emission), is characteristically 
different in the sense that it is a stochastic process.  Starting with a group of partons 
of fixed energy, the bremsstrahlung process  results in a broad spectrum of final partons 
of width comparable to its mean energy loss.  Further, the LPM effect has different 
parametric dependence on energy for soft and hard parts of emitted gluon spectrum. 
As  compared to the case of bremsstrahlung photon emission, the gluon emission 
also involves an  enhancement mechanism  when the emitted gluon and quark are nearly 
collinear, thereby a need to consider ladder diagrams \cite{amy3}.   
However, unlike the emission of electromagnetic quanta, the  emitted 
hard gluon  feels  the random colored background field. The resummation 
of these ladder diagrams leads to Swinger-Dyson type integral equations. 
In this work, we  follow the formalism given in \cite{amy3}-\cite{moore2} 
which implements  LPM effects by resumming the ladder diagrams. 
For calculating parton energy loss  arising from gluon radiation, one needs the differential gluon emission rates
and this is given  by  Eq.\ref{lpm}  \cite{moore2} in terms of the {\bf F}({\bf h},p,k)  
function. The bremsstrahlung integral equations  determine the 
gluon emission amplitude {\bf F}({\bf h},p,k) function given by  
Eq.\ref{amyeq}. Here, the two dimensional vector $\bf{h}=p\times k$ 
is of the order of $O(gT^2)$. It  is a measure of collinearity and its  
magnitude is small compared to $\bf{p k}$. The term $\delta {E}({\bf{h}},p,k)$ in 
Eq.\ref{amyeq} is the energy differential between initial and final states. Here, 
$m_k^2=m_D^2/2=2g^2T^2/3$ and other quark thermal masses are $m^2=g^2T^2/3$. 
This formalism is very similar to the photon emission integrals, however,
as mentioned before, the emitted gluon has color and therefore  interacts  
with other scattering centers as well as soft background fields \cite{amy3}. 
Accordingly, there are three terms in the integral equations of \ref{amyeq} 
involving collision kernel $C({\bf{h}})$. A typical  ladder diagram for gluon 
emission is shown in Figure \ref{ladderg}. 
We solve this  integral equation for {\bf F}({\bf h},p,k) function, 
by using   iterations method as discussed in \cite{surya1}. 
The  {\bf F}({\bf h},p,k) distributions for various values of 
parton and gluon momenta (p,k) were obtained considering the 
mechanism  $g\rightarrow gg$,  $q\rightarrow gq$ and $g\rightarrow q\bar{q}$ processes 
using  relevant factors ($N_s = 2,d_s = d_A(=8)$, and  $C_s = C_A(=3),C_F=4/3$).
Fig.\ref{gluonpdist1} shows  a few  ${\bf F}({\bf h},p,k)$ distributions for 
these two processes (for high parton momenta) for various values of 
 parton and gluon momenta (p,k). The real part is shown in figure (a) and negative 
of imaginary part in figure (b). Fig.\ref{gluonpdist1}(c)  shows  the real 
part of the distributions for $g-gg$ process for high and low  values of  
incoming  and outgoing gluon momenta  (p,k).
The calculations for  $g\rightarrow q\bar{q}$  process are shown in Fig.\ref{gluonpdist2}.
In all these figures, $p$ always stands for the incoming parton momentum.
\begin{eqnarray}
\frac{d\Gamma_{g}^{LPM}}{d^3{\bf{k}}} &=& \frac{\alpha_s}{4\pi^2k^2}\sum_s N_sd_sC_s   \int_{-\infty}^{+\infty}\frac{dp}{2\pi}\int \frac{d^2{\bf h}}{4\pi^2}\nonumber\\ 
&& n_s(p+k) [1\mp n_s(p)][1+n_b(k)] \nonumber\\
&& \frac{1}{k^3}|  \Gamma^s_{p\rightarrow p+k}|^2 2{\bf{h}}\cdot \Re {\bf{F}}_s({\bf{h}};p,k) ~~~~~
\label{lpm}
\end{eqnarray}
\begin{eqnarray}
2{\bf{h}}&=& {i}\delta {E}({\bf{h}},p,k){\bf{F(h)}} + g^2\int\frac{d^2{\bf{h}}}{(2\pi)^2}C({\bf{h}}) \nonumber\\
  &&  \times \left\{(C_s-C_A/2)\left[{\bf{F(h)}}-{\bf{F}}({{\bf{h}}}-k{\bf{h}})\right]\right. \nonumber \\
&&\left. + (C_A/2) \left[{\bf{F(h)}}-{\bf{F}}({{\bf{h}}}+p{\bf{h}})\right]\right. \nonumber \\
 && \left.+ (C_A/2)\left[{\bf{F(h)}}-{\bf{F}}({{\bf{h}}}-(p-k){\bf{h}})\right] \right\}~~~~~ \label{amyeq}
\end{eqnarray}
\begin{eqnarray}
\delta {E}({\bf{h}},p,k)&=&\frac{{\bf{h^2}}}{2pk(p-k)} \nonumber\\
&& + \left[  \frac{m_k^2}{2k} + \frac{m_{p-k}^2}{2(p-k)} - \frac{m_p^2}{2p}\right] ~~~~~~~~~ \label{deltae}\\
{C}({\bf {h}})&=&\frac{1}{\bf {h}^2}-\frac{1}{({\bf {h}^2 }+1)}
\label{kernel}
\end{eqnarray}
\section{Generalized Emission Functions for gluon emission}
\begin{figure}
\includegraphics[height=8.cm,width=8.0cm]{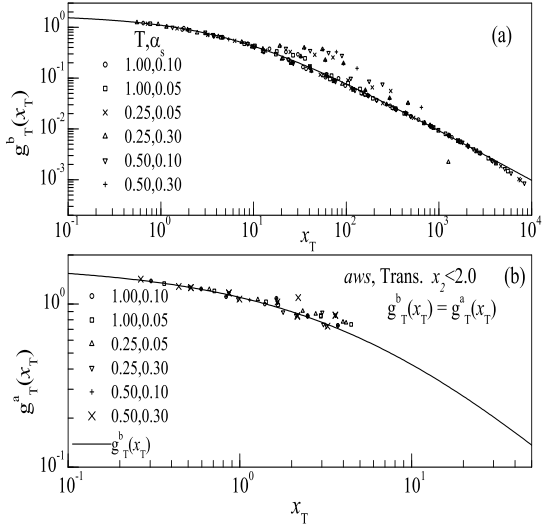}
\caption{  (a)~ The  dimensionless  emission  function  $g^b_T(x)$  versus dynamical variable $x_T$ defined in Eq.\ref{xt}. 
 Six cases of  temperature and coupling constant values  considered are mentioned in figure labels in different 
type symbols. The symbols represent the integrated values of ${\bf p_\perp}$ distributions as a function  
of $\{p_0,q_0,Q^2,T,\alpha_s\}$ values. The solid curve is an 
empirical fit given by Eq.\ref{btgxemp}.}
\label{btatgx}
\end{figure}

Before we discuss the emission functions for gluon emission, it is very
instructive to recall the emission functions for photon emission.
In our previous works \cite{surya1,svsprc}, we showed generalized 
photon emission function by integrating (Eq.\ref{itdef}) the corresponding
 ${\bf p_\perp{\cdot}f(p_\perp)}$ distributions (see \cite{surya1}.  
\begin{eqnarray}
I_{T}&=&\int \frac{\bf d^2\tilde{p}_\perp}{(2\pi)^2}  {\bf \tilde{p}_\perp{\cdot}\Re\tilde{f}(\tilde{p}_\perp)} \label{itdef}
\end{eqnarray}
\begin{eqnarray}
x_0&=&\frac{|(p_0+q_0)p_0|}{q_0T  } \label{x0} \\
x_1 &=&x_0\frac{M_\infty^2}{m_D^2}  \label{x1}\\
x_T&=&x_1+x_2   \label{xt} 
\end{eqnarray}
 \begin{eqnarray}
g(x)&=&I \times c  \label{gbatl} \\
c&=&\frac{1}{x_1^2}    \label{cbt}
\end{eqnarray}
\begin{eqnarray}
g^b_T(x)&=&\frac{10.0}{5.6+2.5\sqrt{x}+x}  \label{btgxemp}
\end{eqnarray}
\begin{eqnarray}
\Im{\Pi^\mu}_{R\mu} &=& \frac{e^2N_c}{2\pi} \int_{-\infty}^\infty  dp_0 [n_F(r_0)-n_F(p_0)] \otimes \nonumber \\
&&\left(Tm_D^2\right) \left[ \frac{p_0^2+r_0^2}{2(p_0r_0)^2}\left(\frac{g^i_T \left(x_T\right)}{c^i_T}\right)+ \right. \nonumber \\
&& \left. \frac{1}{\sqrt{\left|p_0r_0\right|}}\frac{Q^2}{q^2} \left(\frac{1}{m_D}\right) \left(\frac{g^i_L \left(x_L\right)}{c^i_L}\right)  \right] 
\label{impolargx}
\end{eqnarray} 
$I$ are in general functions of  \{$p_0,q_0,Q^2,T,\alpha_s$\} and therefore,
we defined the generalized emission functions (GEF) $g$ in Eq.\ref{gbatl},
which  are functions of only $x$  variables. These  GEF ($g$)  are obtained 
from corresponding $I$ values by multiplying with $c$ coefficient functions given 
in \cite{surya1}. As an example, in Figure \ref{btatgx}, we show the results 
for  GEF for bremsstrahlung (Fig.\ref{btatgx}(a)).
The solid curve in (a) is the empirical fit to this emission function, 
given by Eq.\ref{btgxemp}.  Consequent to finding the emission functions like 
those  given by Eq.\ref{btgxemp}, we  expressed the imaginary part of 
retarded photon polarisation tensor for any $p_0,q_0,Q^2,T,\alpha_s$  
values by using Eq.\ref{impolargx}. Using this approach, we obtained 
the phenomenological fits to virtual photon emission rates from
QGP for ladder  processes with LPM effects \cite{surya1}.  We provided
simple  phenomenological  formulae  which are useful 
in  model calculations for experimental dilepton yields. 
\begin{figure}
\includegraphics[height=12.0cm,width=8.0cm]{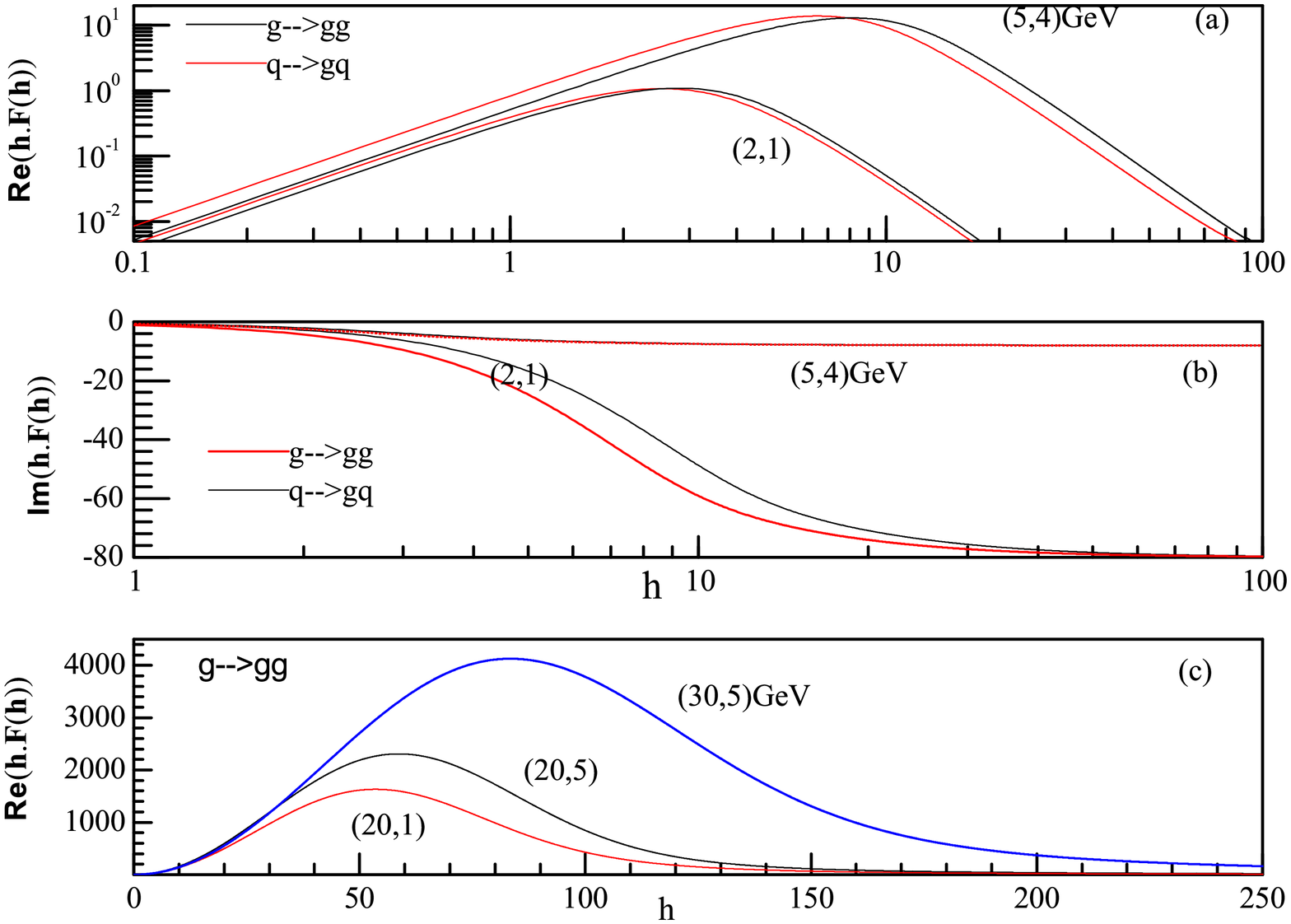}
\caption{ (a)~Shows the  real part of   ${\bf h}$ distributions 
of  the ${\bf h} \cdot{\bf  \Re{F}}({\bf  h})$ for gluons. 
Various curves are for various parton momenta (p) and gluon momenta (k)  
values as mentioned in figure labels as (p,k). The distributions are 
obtained using iterations method.\\
(b)~Shows the imaginary part of   ${\bf h}$ distributions 
of the ${\bf h}\cdot{\bf \Im{F}}({\bf h})$ \\
(c)~Shows the  real part of   ${\bf h}$ distributions 
of  the ${\bf h} \cdot{\bf  \Re{F}}({\bf  h})$ for pure glue process. 
Various curves are for various parton momenta (p) and gluon momenta (k)  
values as mentioned in figure labels as (p,k).
\label{gluonpdist1}
}\end{figure}
\begin{figure}
\includegraphics[height=12.cm,width=8.0cm]{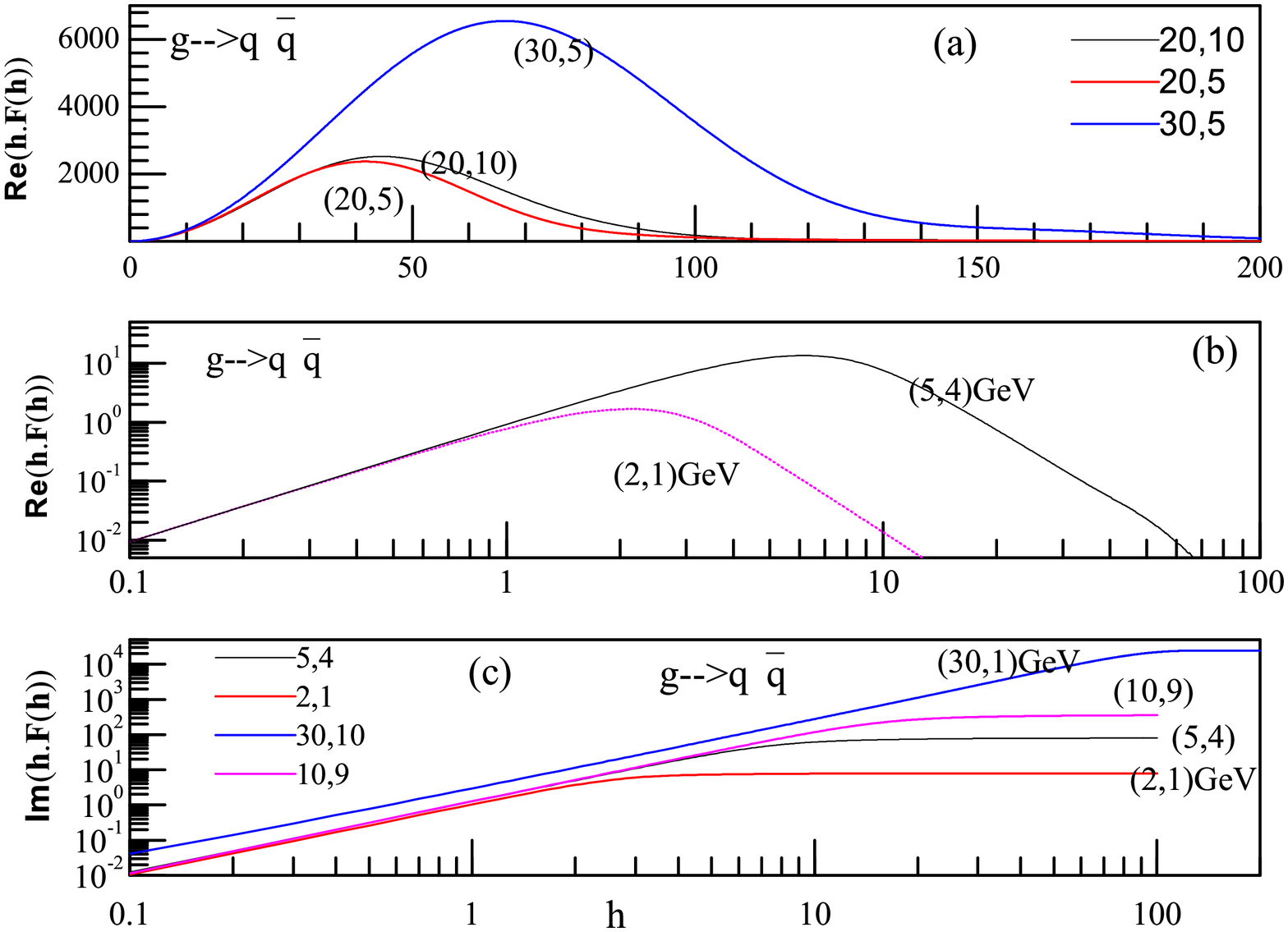}
\caption{ (a)~Shows the  real part of   ${\bf h}$ distributions 
of  the ${\bf h} \cdot{\bf  \Re{F}}({\bf  h})$ for $g\rightarrow q\bar{q}$ process. 
Various curves are for various gluon momenta (p) and quark momenta (k)  
values as mentioned in figure labels as (p,k). The distributions are 
obtained using iterations method.\\
(b)~Shows the  real part of   ${\bf h}$ distributions 
of  the ${\bf h} \cdot{\bf  \Re{F}}({\bf  h})$ for $g\rightarrow q\bar{q}$ process. 
Various curves are for various gluon momenta (p) and quark momenta (k)  
values as mentioned in figure labels as (p,k).\\
(c)~Shows the imaginary part of   ${\bf h}$ distributions 
of the ${\bf h}\cdot{\bf \Im{F}}({\bf h})$ 
\label{gluonpdist2}
}\end{figure}
\begin{figure}
\includegraphics[height=8.0cm,width=8.0cm]{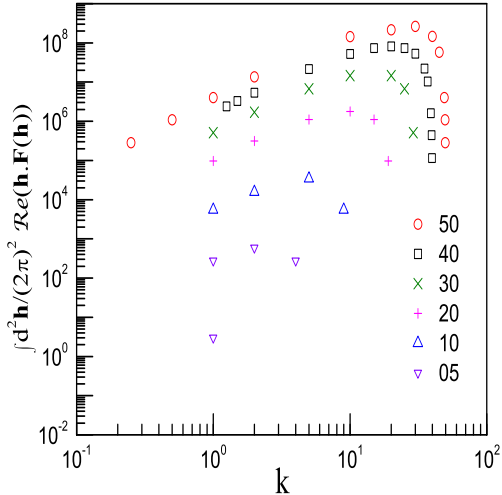}
\vspace{-0.5cm}
\caption{ ~The integral of gluon {\bf h}  distributions shown in
 previous figures for the process 
$g\rightarrow gg$.   $ \int \frac{\bf d^2{h}}{(2\pi)^2}{\bf h} \cdot{\bf  \Re{F}}({\bf  h})$ versus
emitted gluon momentum  (k). Here incoming parton momenta (in the present case, 
gluon momenta p) has not yet been integrated.  The different parton  momentum (gluon $p$) values are 
shown on the  figure and temperature of plasma is taken T=1.0GeV.
\label{g2gghfintkdist}
}\end{figure}
\begin{figure}
\includegraphics[height=10.0cm,width=8.0cm]{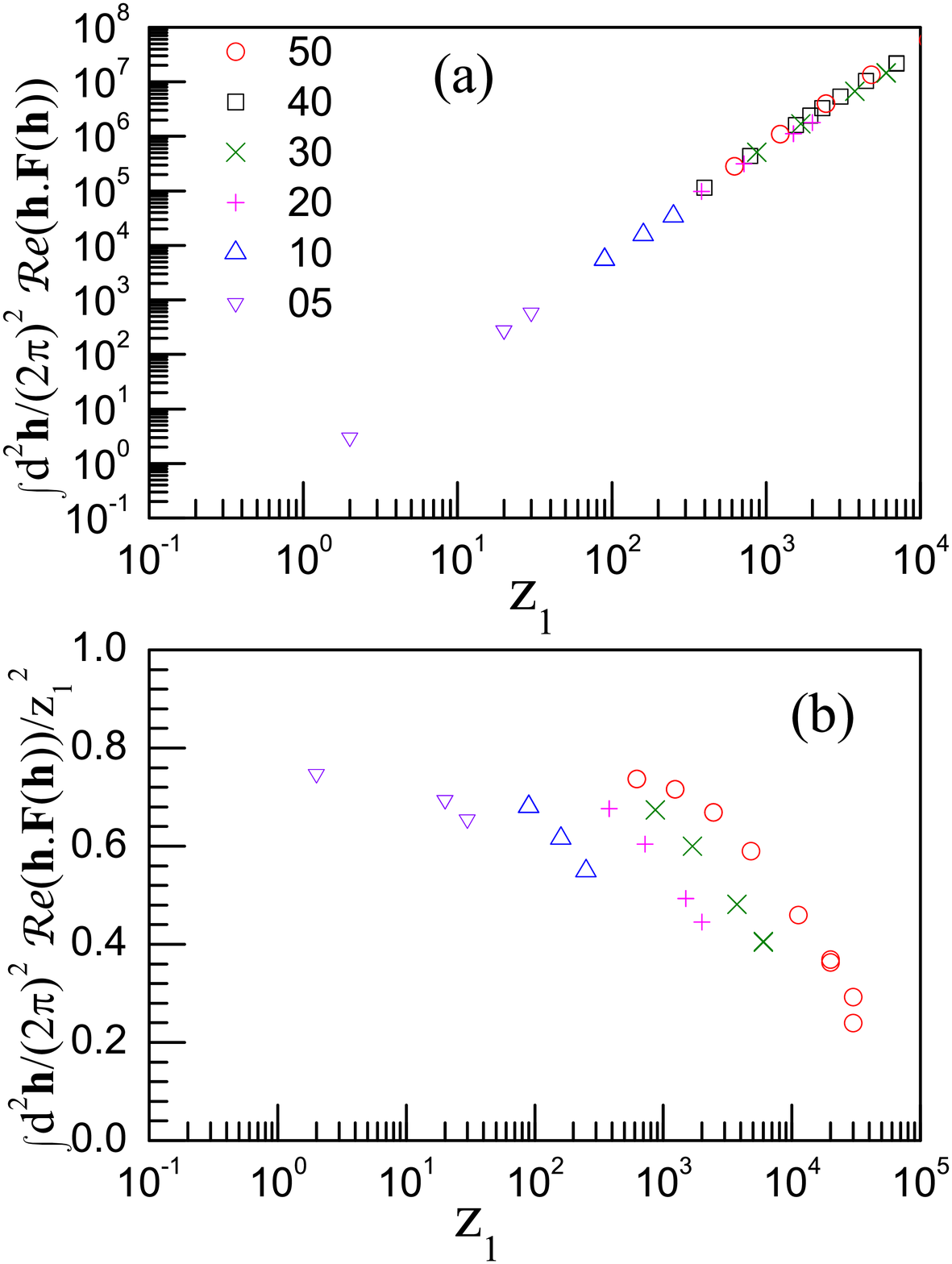}
\caption{ ~The integral of gluon {\bf h} distributions  for the process 
$g\rightarrow gg$.  The plot shows  
$\int \frac{\bf d^2{h}}{(2\pi)^2} {\bf h} \cdot{\bf  \Re{F}}({\bf  h})$ versus
the variable  z$_1$. The different parton  momentum (gluon $p$) values are 
shown on the  figure and temperature of plasma is taken T=1.0GeV.
\label{g2gghfintz1dist}
}\end{figure}
Following the procedure of generalized photon emission function, we now try to obtain 
the generalized gluon emission functions. For this, we integrate these distributions,
that have been shown in Figs.\ref{gluonpdist1},\ref{gluonpdist2} over the variable $\bf h$, 
as given in Eq.\ref{igdef}. This quantity $I$ is strongly process dependent and is 
a function incoming and outgoing parton momenta, plasma temperature and strong 
coupling strength as denoted by the variables $p,k,T$ and $\alpha_s$. The integrated quantity $I$
is plotted versus $k$ in Figure \ref{g2gghfintkdist} for the process $g\rightarrow gg$. 
Figure shows $I$ plotted for different values of $p$ labeled on the figure. 
As seen in figure, the $I$ values are scattered. Therefore, we defined a variable $z_1$ as given 
in Eq.\ref{z1def}. We show $I$  values versus $z_1$ variable in figure \ref{g2gghfintz1dist}.
As seen in figure (a), it exhibits a linear behavior on log-log plot, extending over  nine orders of magnitude.
This apparently gives an impression that $z_1$ is a good dynamical scale for the process $g\rightarrow gg$.
In order to examine this, we plot  $I/z_1^2$ in figure (b).  As seen in figure (b), the $I/z_1^2$ values are 
scattered and exhibit no useful trends, showing that $z_1$ is not a dynamical variable for this 
process. Therefore, we now define the dynamical variable $x$ and a function $g$,
 for $g\rightarrow gg$ process as given in Eqs.\ref{xdef},\ref{gdef}. In the Fig.\ref{gxg2gg}, we show the function $g$
versus $x$. As seen in figure, all $I$ values for different parton momenta merge. We fit this data with an empirical 
curve together with parameters as given in Eqs.\ref{gxg2gg},\ref{gxg2ggpar}. We denote this function in Eq.\ref{gxg2gg} as 
gluon emission function for the process $g\rightarrow gg$. \\

\begin{figure}
\includegraphics[height=8.0cm,width=8.0cm]{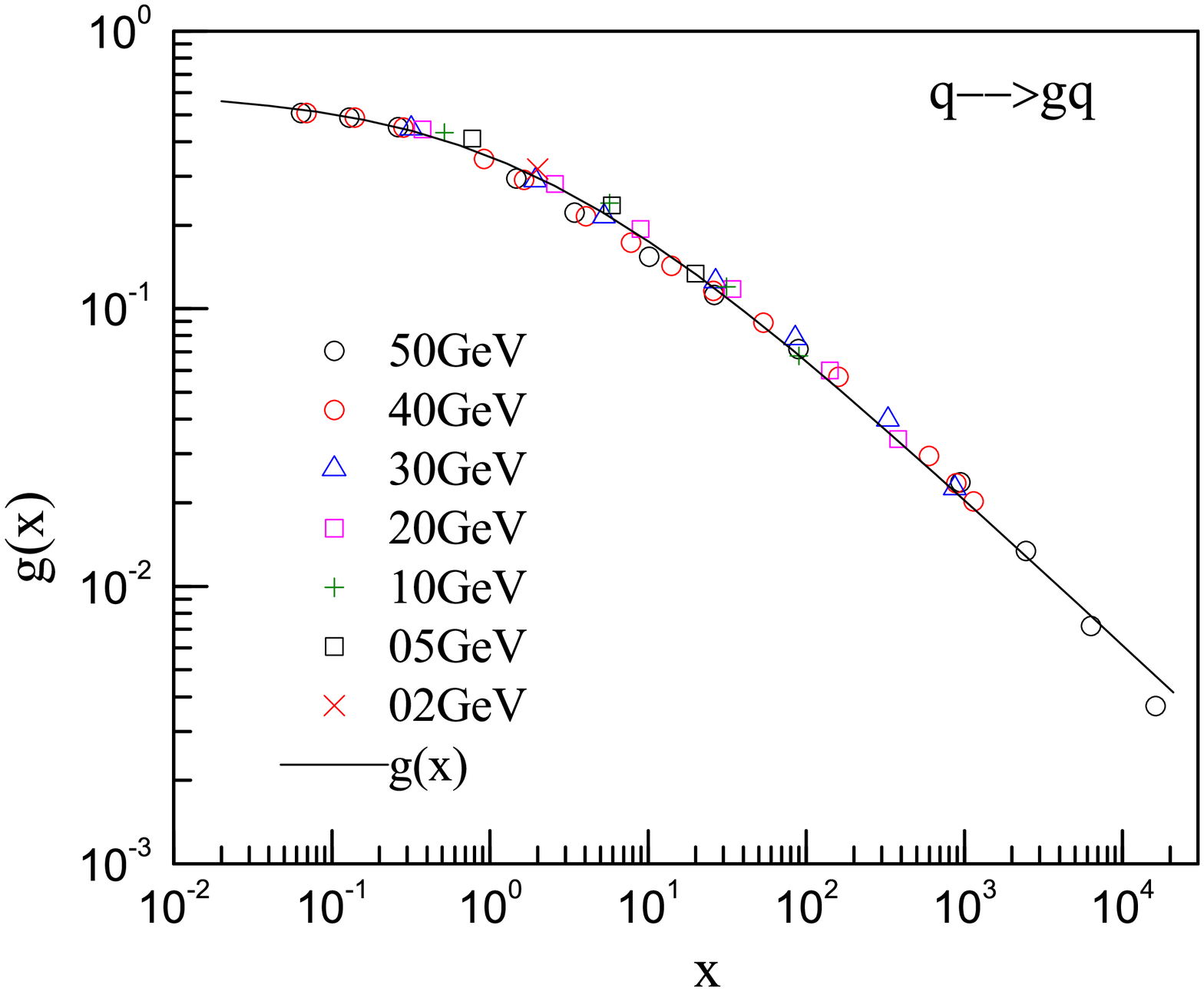}
\vspace{-0.5cm}
\caption{ ~The integral of gluon {\bf h} distributions for the  process 
$q\rightarrow gq$.  This plot shows
 $\int \frac{\bf d^2{h}}{(2\pi)^2}{\bf h}\cdot{\bf \Re{F}}({\bf  h})$ versus
the new variable  $x$. As before, incoming parton  momentum (quark $p$) values are 
shown on the  figure and temperature of plasma is taken T=1.0GeV.
\label{gxq2gq}
}\end{figure}

\begin{figure}[htb]
\includegraphics[height=8.0cm,width=8.0cm]{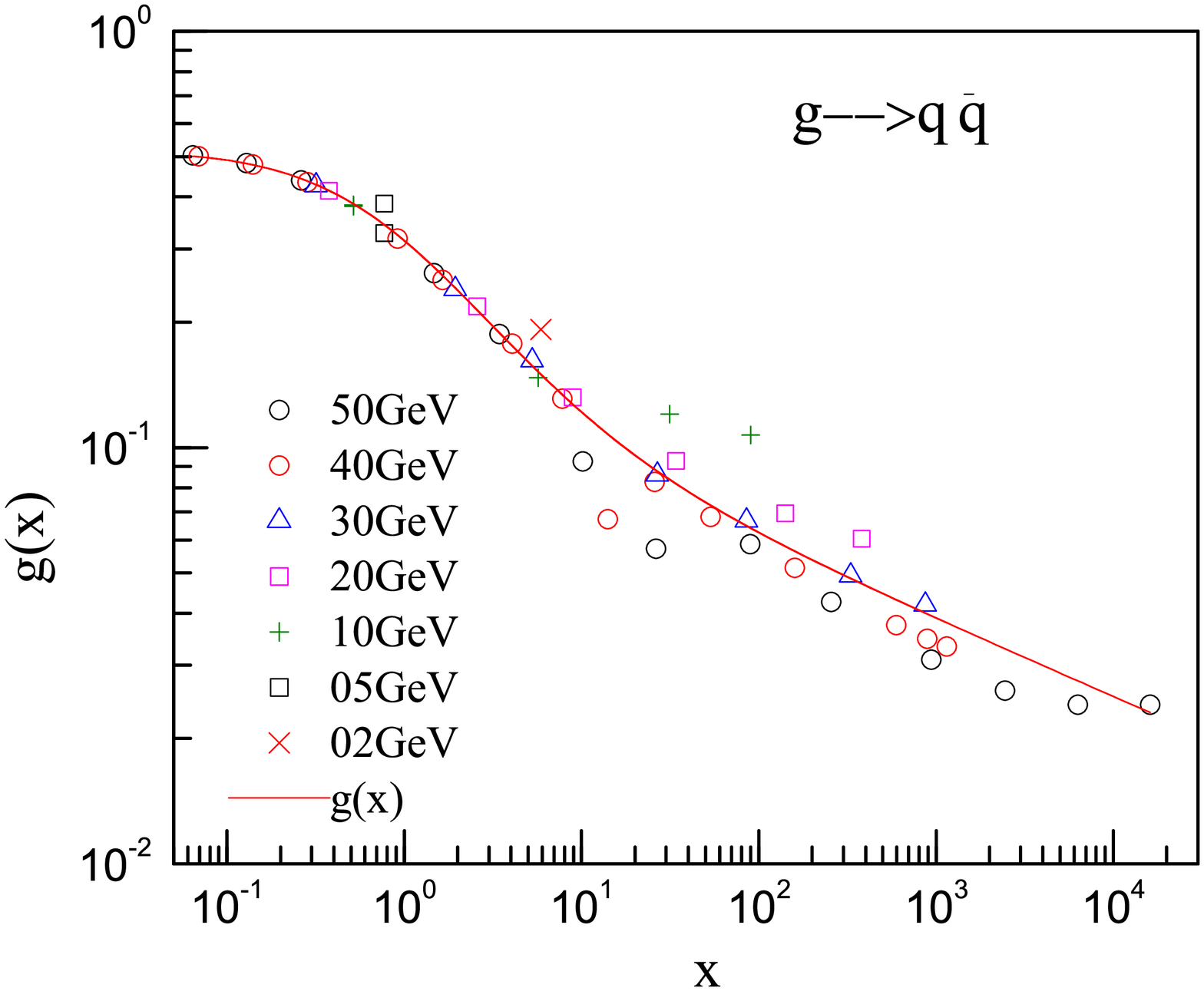}
\vspace{-0.5cm}
\caption{ ~The integral of gluon {\bf h} distributions for the  process 
$g\rightarrow q\bar{q}$.  This plot shows
 $\int \frac{\bf d^2{h}}{(2\pi)^2}{\bf h}\cdot{\bf  \Re{F}}({\bf  h})$ versus
the new variable  $x$. The different incoming parton  momentum (gluon $p$) values are 
shown on the  figure and temperature of plasma is taken T=1.0GeV.
\label{gxg2qqbar}
}\end{figure}

\begin{figure}
\includegraphics[height=4.0cm,width=8.0cm]{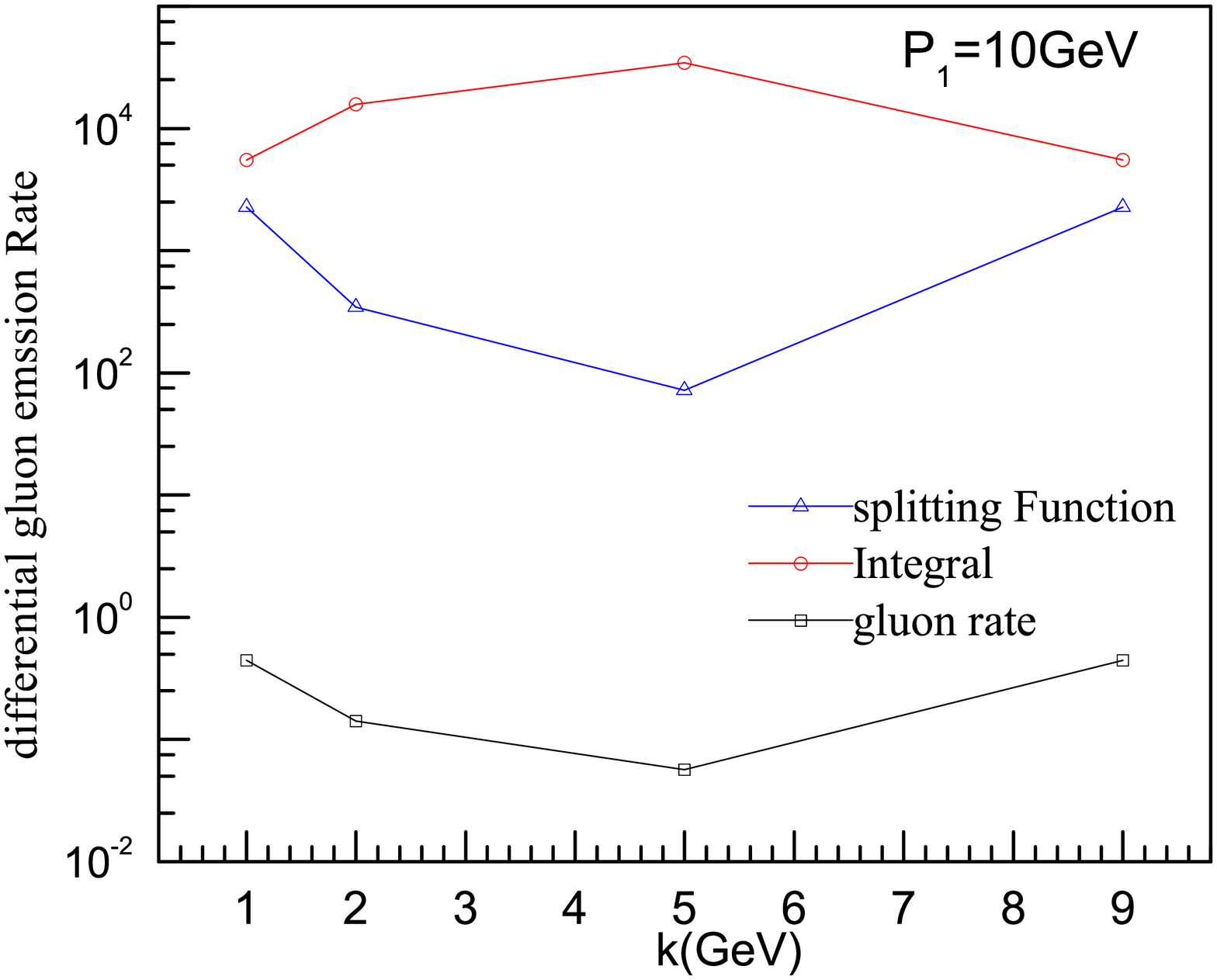}
\includegraphics[height=4.0cm,width=8.0cm]{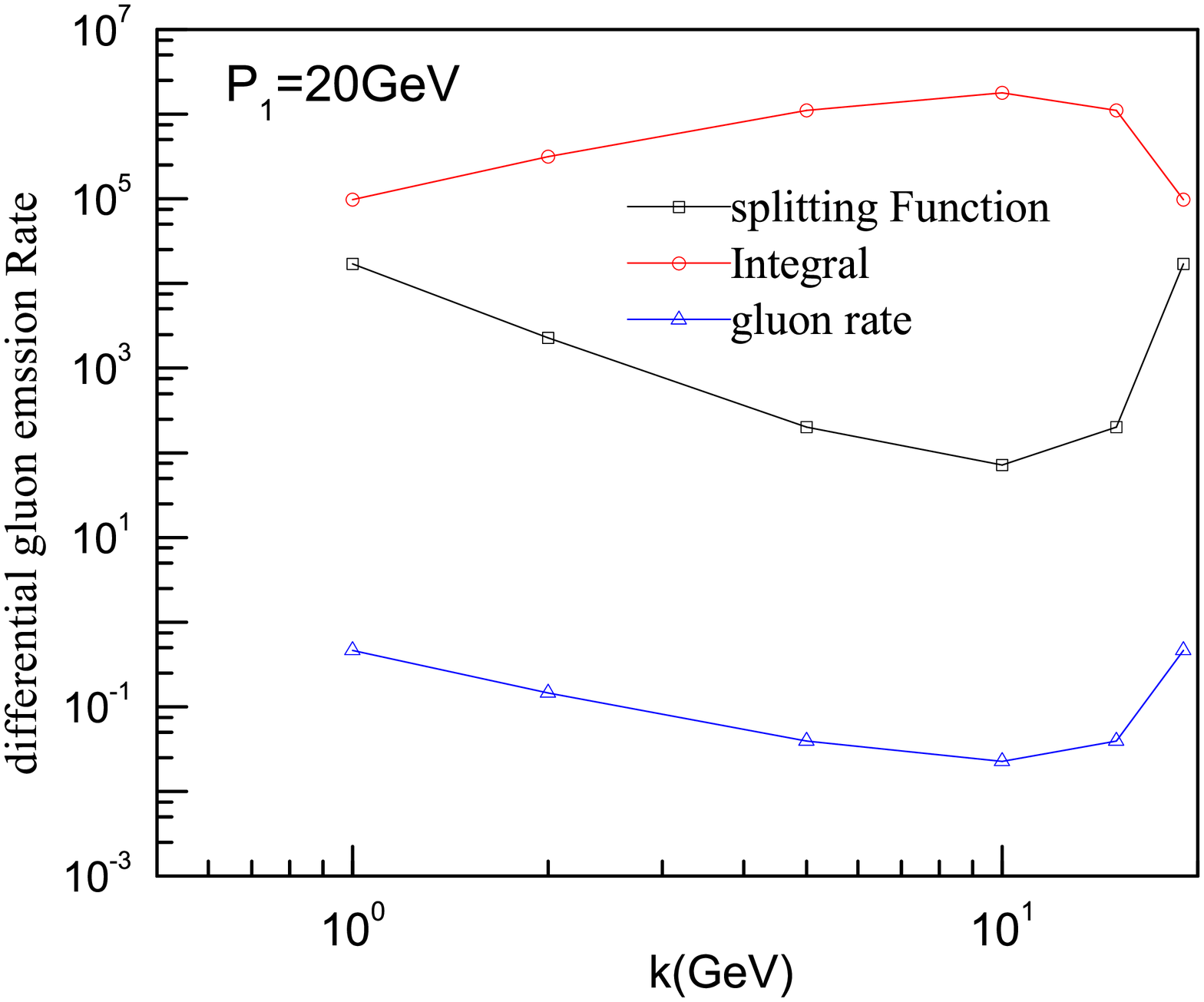}
\includegraphics[height=4.0cm,width=8.0cm]{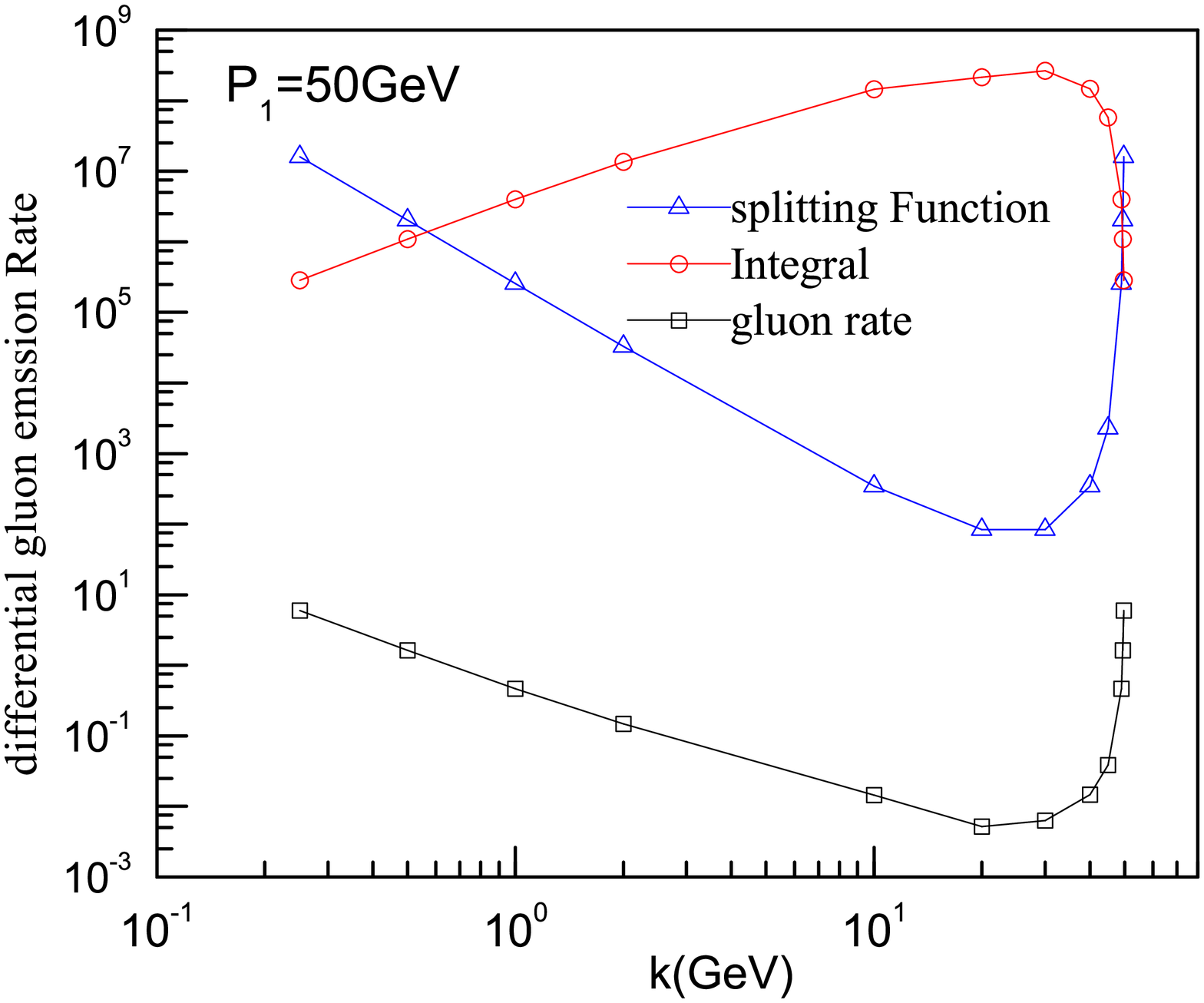}
\vspace{-0.5cm}
\caption{ ~ Black line versus gluon momentum (k) shows the integrand (without p integration)  of the 
differential gluon emission rate of Eq.\ref{lpm} $\frac{d\Gamma_{g}^{LPM}}{d^3{\bf{k}}} $ for the process
$g\rightarrow gg$. The blue curve represents the splitting function for $g\rightarrow gg$ given 
by $|  \Gamma^s_{p\rightarrow p+k}| $ of Eq.\ref{lpm} and the red curve represents the integral 
value $\int \frac{d^2{\bf h}}{(2\pi)^2} 2{\bf{h}}\cdot \Re {\bf{F}}_s({\bf{h}};p,k) $. The 
incoming parton (gluon in this case) momentum $p$ (labeled as $p_1$ in figure) for this figure is fixed $p=10,20,50$GeV as indicated on the figure.
Temperature of plasma is T=1.0GeV.
\label{g2ggrate}
}\end{figure}

\begin{figure}
\includegraphics[height=4.0cm,width=8.0cm]{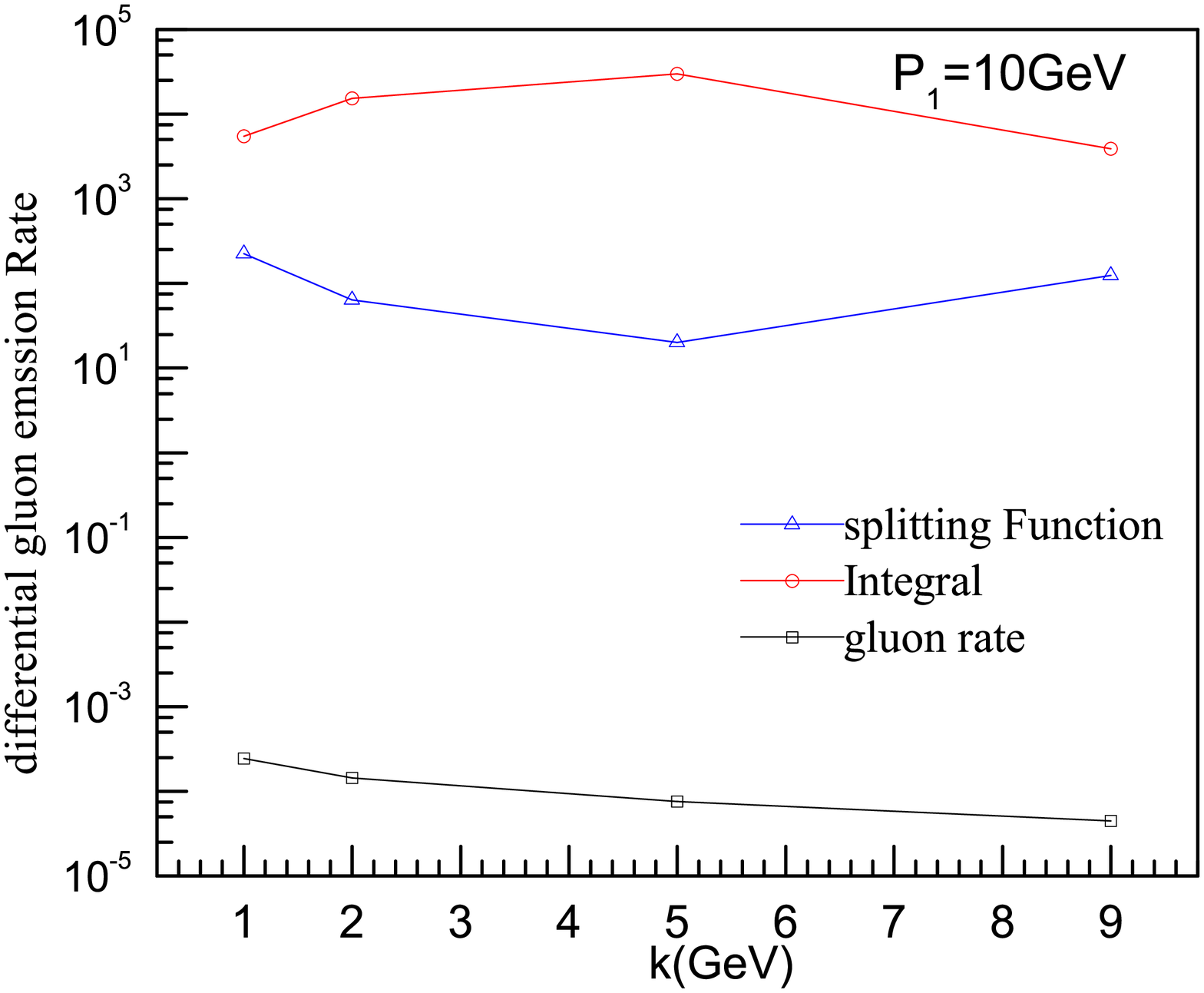}
\includegraphics[height=4.0cm,width=8.0cm]{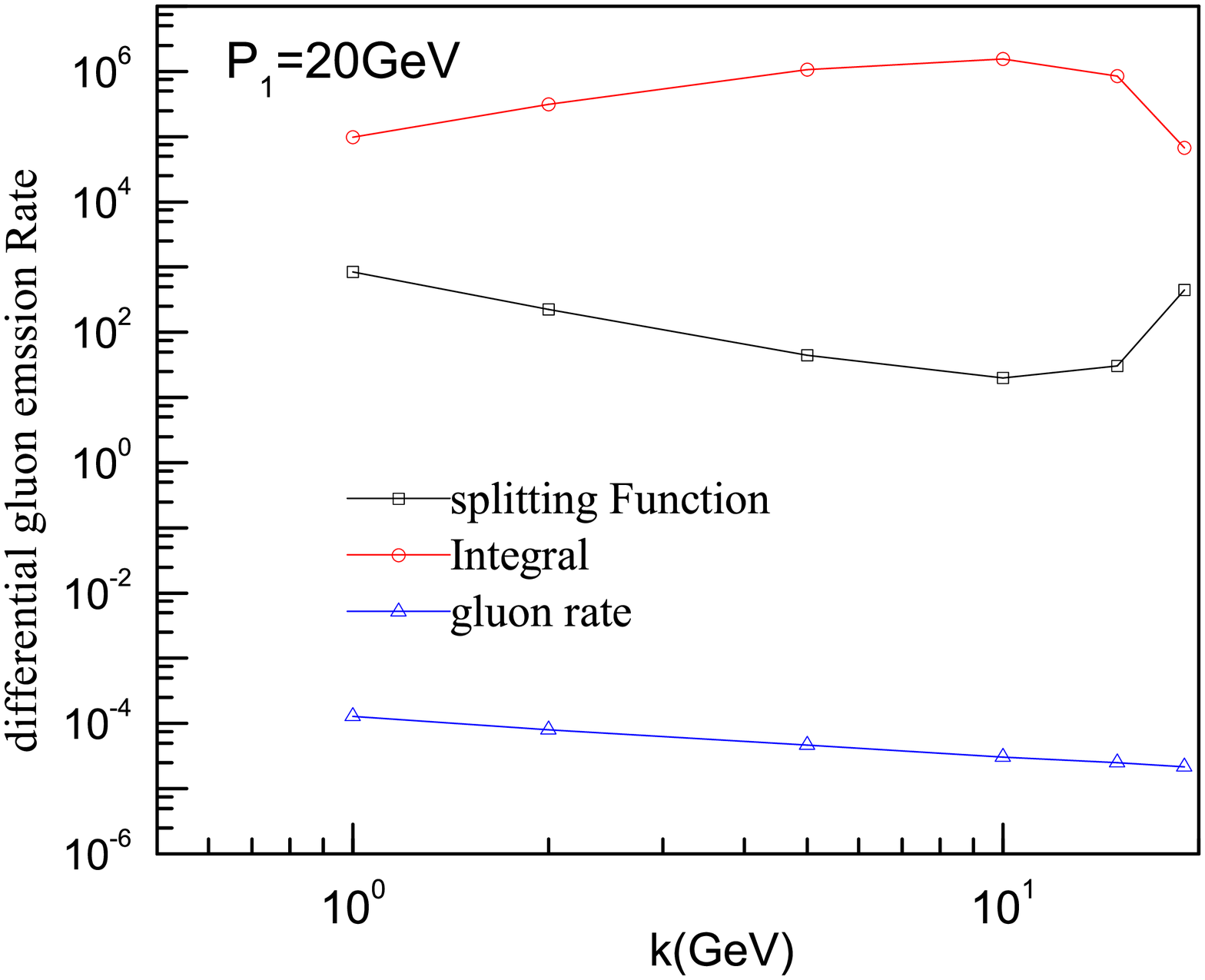}
\includegraphics[height=4.0cm,width=8.0cm]{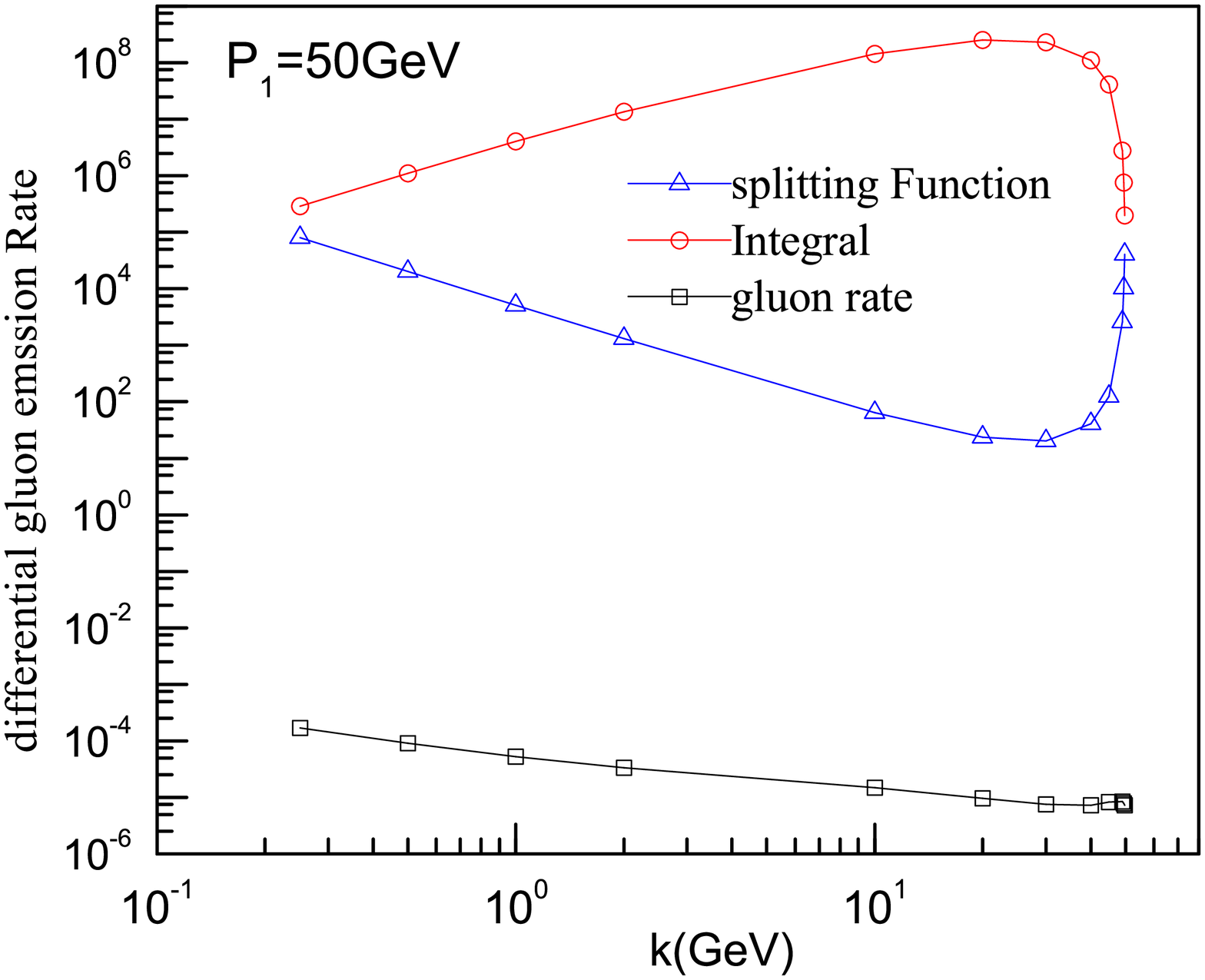}
\caption{ ~ Black line versus gluon momentum (k) shows the integrand of the differential 
gluon emission rate of Eq.\ref{lpm} $\frac{d\Gamma_{g}^{LPM}}{d^3{\bf{k}}} $ for the process
$q\rightarrow gq$. The curves are as in Figure \ref{g2ggrate}.
\label{q2gqrate}
}\end{figure}

\begin{figure}
\includegraphics[height=3.50cm,width=8.0cm]{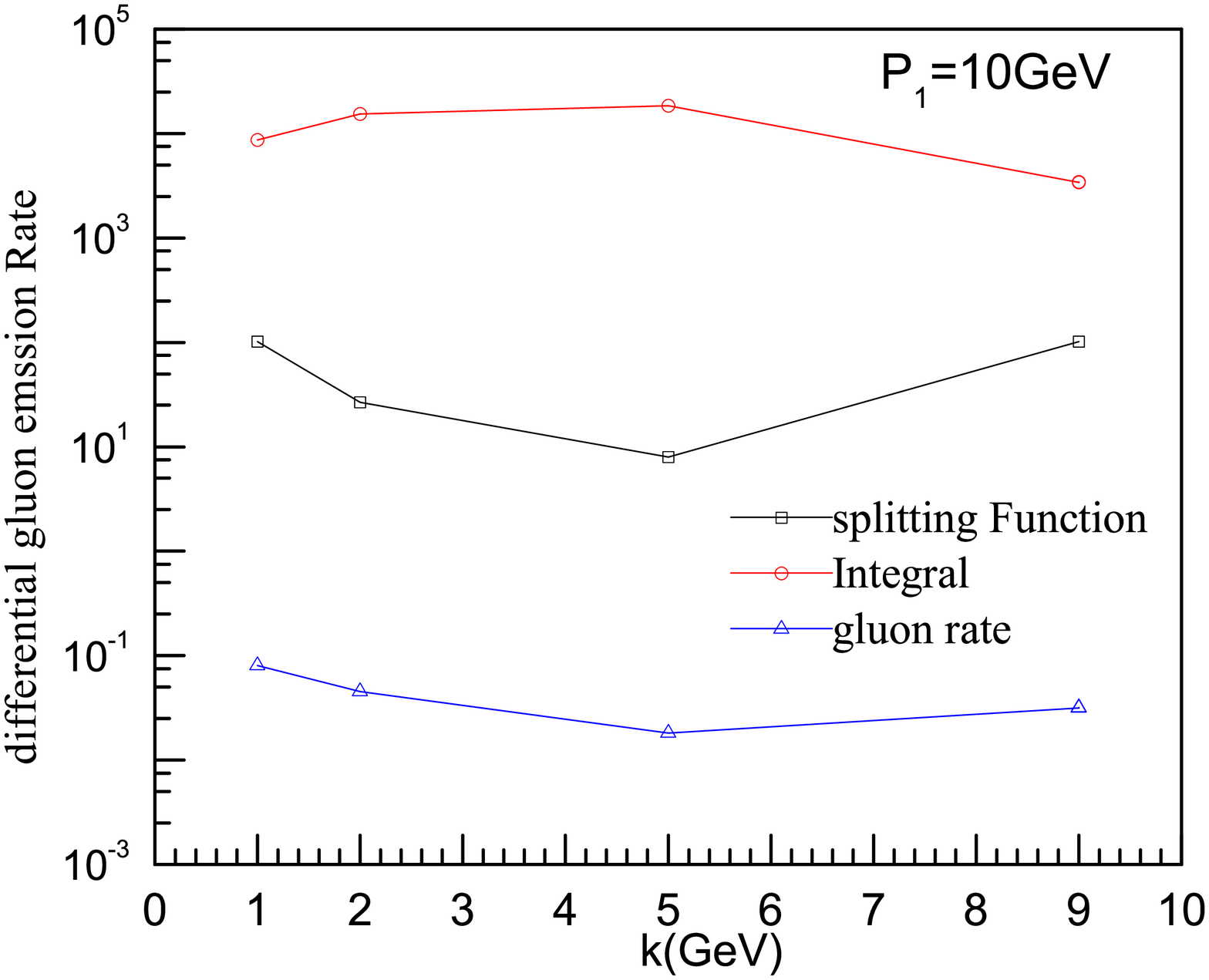}
\includegraphics[height=3.50cm,width=8.0cm]{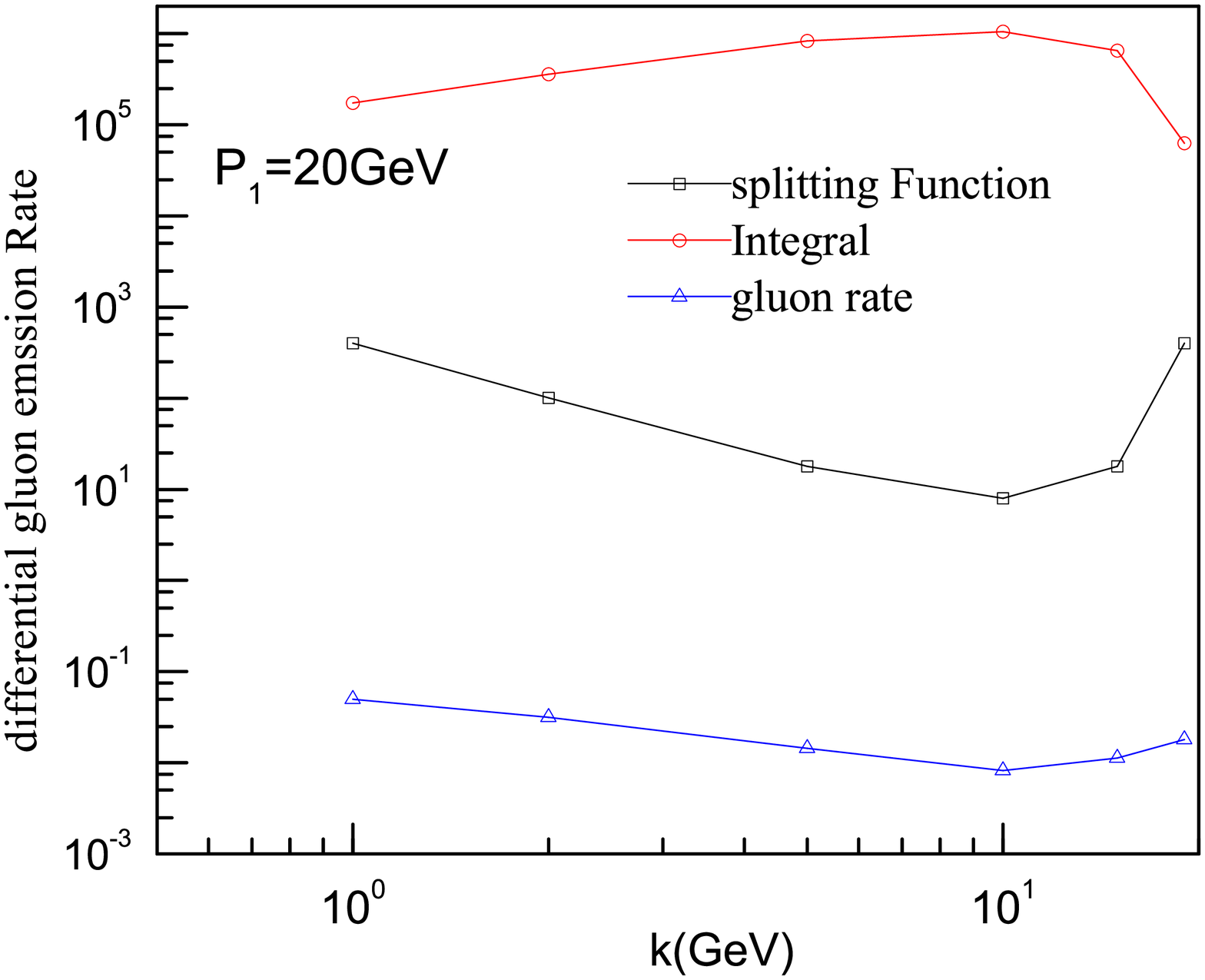}
\includegraphics[height=3.50cm,width=8.0cm]{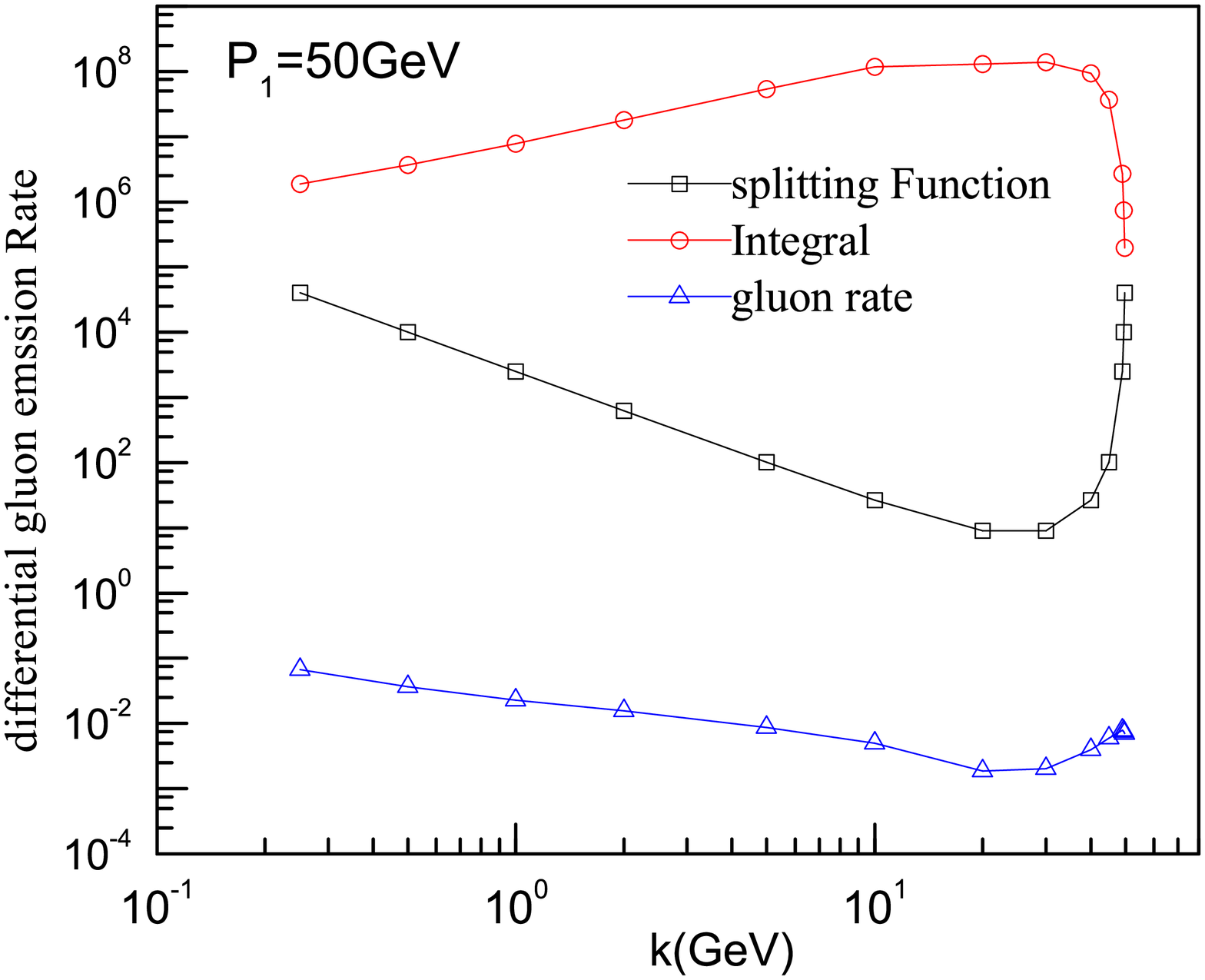}
\caption{ ~ Black line versus quark momentum (k) shows the integrand of differential 
emission rate of Eq.\ref{lpm} $\frac{d\Gamma_{g}^{LPM}}{d^3{\bf{k}}} $ for the process $g\rightarrow q\bar{q}$. 
The details are as in Figure \ref{g2ggrate}.
\label{g2qqbarrate}
}\end{figure}

\noindent
We carried out this for the processes $q\rightarrow gq$ and $g\rightarrow q\bar{q}$. The $I$ values for 
these two processes also donot exhibit any trends as a function $z_1$ variable, however, $x$ remains a good dynamical 
variable. We show these results for the process $q\rightarrow gq$  in Fig.\ref{gxq2gq} versus $x$. 
The curve in this figure is given by empirical fit and parameters in Eqs.\ref{gxq2gqeq},\ref{gxq2gqpar}.
Therefore, for $q\rightarrow gq$ , we define gluon emission function $g(x)$ as given in Eq.\ref{gxq2gqeq}.
We performed these calculations for the process $g\rightarrow q\bar{q}$ and these results are shown 
in Fig.\ref{gxg2qqbar}. The curve in this figure is given by empirical fit and parameters 
 in Eq.\ref{gxg2qqbareq},\ref{gxg2qqbarpar}. Therefore, for the process  $g\rightarrow q\bar{q}$ , 
we define gluon emission function $g(x)$ as given in Eq.\ref{gxg2qqbareq}. \\
\noindent
After obtaining the gluon emission function $g(x)$ for these three processes, we can perform 
$p$ integrations required in Eq.\ref{lpm}, i.e. integration in terms of dynamical variables $x$.
This will give us differential gluon emission rates. In the jet-quenching studies, one needs to 
estimate the  energy loss of high energy partons while traversing the QGP medium. In this 
problem, the differential gluon emission rates estimated by integrating over $p$ variable, will 
have to be coupled in order to determine the differential energy loss. These results were already shown by
\cite{amy3,moore1,moore2,mazum}. \\
In the following, we examine the integrand of differential gluon emission rate of the Eq.\ref{lpm}, 
{\it i.e.,}  without $p$ integration, and we denote this in short by gluon rate, 
which should not be confused with the gluon emission rates after integrating over $p$.  
We obtained the $I$ values by integrating the $h$ distributions 
$\int \frac{\bf d^2{h}}{(2\pi)^2} {\bf h} \cdot{\bf  \Re{F}}({\bf  h})$.
The splitting function $\left|\Gamma^s_{p\rightarrow p+k}\right|^2 $ are as given 
in \cite{moore2}. In the Fig.\ref{g2ggrate}, we show the $I$ values (red curves), splitting 
function (blue curves) and the gluon rate (black curves) at three different $p$ values
for the process $g\rightarrow gg$.  We show similar results for the process $q\rightarrow gq$
in Fig.\ref{q2gqrate} and for the process $g\rightarrow q\bar{q}$ in Fig.\ref{g2qqbarrate}.\\
\begin{figure}
\includegraphics[height=8.0cm,width=8.0cm]{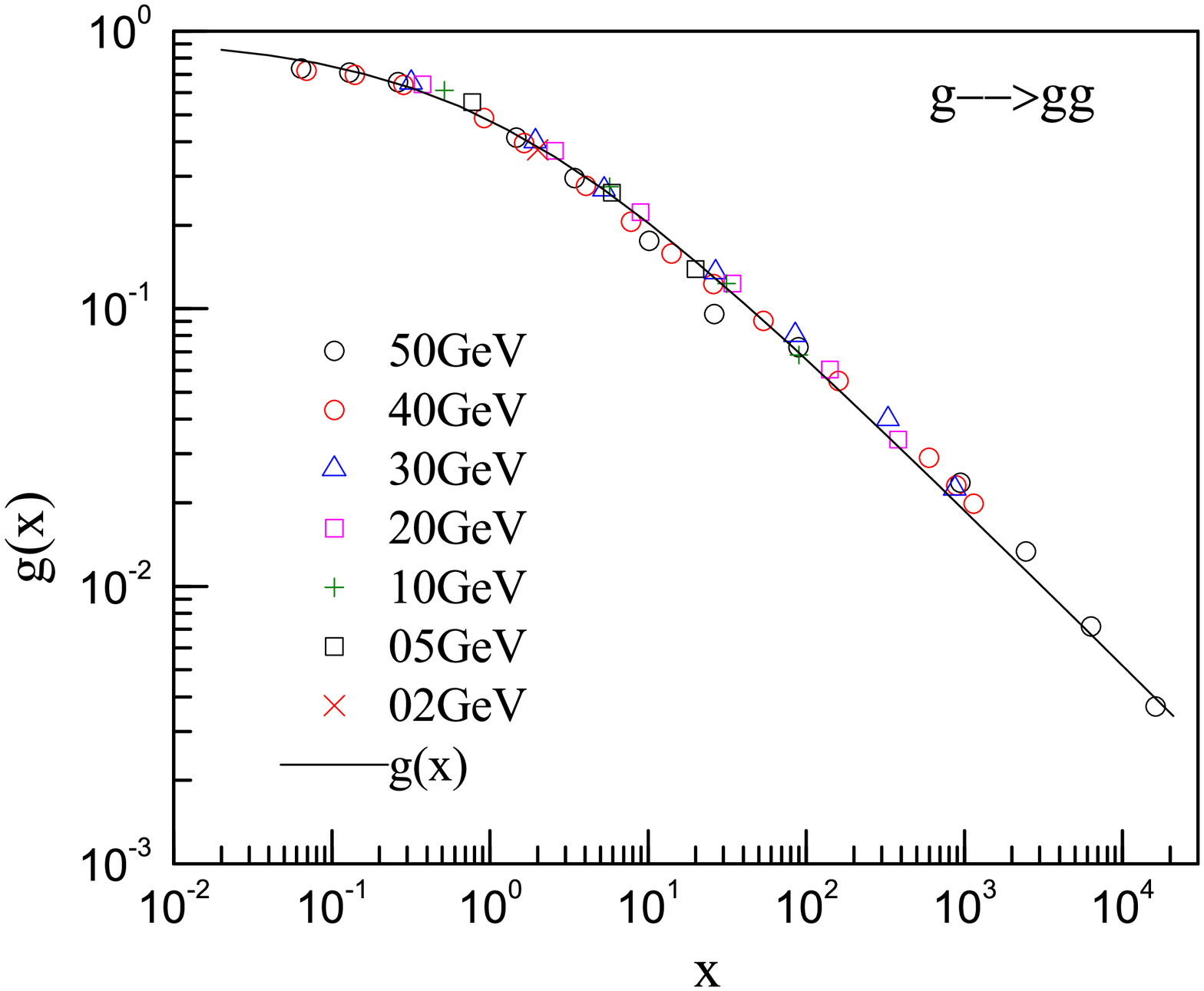}
\vspace{-0.5cm}
\caption{ ~The integral of gluon ${\bf h}$ distributions for the  process $g\rightarrow gg$.  
The plot shows  $\int \frac{\bf d^2{h}}{(2\pi)^2} {\bf h} \cdot{\bf  \Re{F}}({\bf  h})$ 
versus the new variable  $x$. As before, different incoming parton  momentum (gluon $p$) values are 
shown on the  figure and temperature of plasma is taken T=1.0GeV.
\label{gxggg}
}\end{figure}
\begin{eqnarray}
I & = & \int \frac{\bf d^2{h}}{(2\pi)^2} {\bf h} \cdot{\bf  \Re{F}}({\bf  h}) \label{igdef} \\
z_1 & = & \left|\frac{(p-k)pk}{T}\right| \label{z1def}\\
x   & = & \frac{z_1}{{k^{2.35}}} \label{xdef} \\
g   & = & \frac{I}{z_1^{2}}\left(\frac{k}{p}\right) \label{gdef}
\end{eqnarray}
\begin{eqnarray}
g(x)&=& \frac{0.9500}{(1.000+c x^{w1})} \label{gxg2gg} \\
c &=& 0.995328684841907 \\
w1 &=& 0.565842143207086 \label{gxg2ggpar} 
\end{eqnarray}
\begin{eqnarray}
g(x)&=& \frac{a}{(1.000+c x^{w1})} \label{gxq2gqeq}\\
a &=& 0.609492873517025 \\
c &=& 0.725062591204554 \\
w1 &=& 0.533208076507815 \label{gxq2gqpar} 
\end{eqnarray}
\begin{eqnarray}
g(x)&=& \frac{(a+b x^{w1})}{(1+c x^{w2})} \label{gxg2qqbareq} \\
%
a     & =& 0.51922  \\
b & = & 0.185609  \\
c & = & 1.21859    \\
w1 & = & 0.807358 \\
w2 & = & 1.00605  \label{gxg2qqbarpar}
\end{eqnarray}
\clearpage
\noindent
In conclusion, the  gluon emission in quark  gluon plasma  including   LPM  effects has  been
studied   at a fixed  temperatures and strong coupling strength.
We defined a new  dynamical variable $x$ for gluon emission. Further,  we defined  gluon emission functions (GEF)
denoted by $g(x)$ for the processes $g\rightarrow gg$,~$q\rightarrow gq$ and $g\rightarrow q\bar{q}$. 
We  have obtained empirical fits to these GEF   and  provide the functional forms and parameters 
for all the three processes. We compared the differential gluon emission rates (without p-integration) for these three processes.   \\
In terms of the  GEF,  we may calculate the  differential gluon emission rates for these processes. 
These empirical formulae will be useful in calculations of parton energy loss by medium induced gluon radiation.
\acknowledgements{I am thankful to  Drs. A. K. Mohanty, S. Kailas, R. K. Choudhury, S. Ganesan and H. Naik
for fruitful discussions. I thank S.V. Ramalakshmi for her kind co-operation during this work.}
\noindent

\end{document}